\documentclass[preprint,doublecolumn,12pt]{elsarticle}

\usepackage{lineno,hyperref}
\modulolinenumbers[5]

\usepackage{epsfig}
\usepackage{amsmath}
\usepackage{amssymb}
\usepackage{color}
\usepackage{url}
\usepackage{multirow}
\usepackage{algorithm}
\usepackage{algorithmic}
\usepackage{graphicx}
\usepackage{subfigure}
\usepackage{enumerate}
\usepackage{lineno}

\begin{document}

\title{A GPU Based Memory Optimized Parallel Method For FFT Implementation}

\author[mymainaddress]{Fan Zhang\corref{mycorrespondingauthor}}
\cortext[mycorrespondingauthor]{Corresponding author}
\ead{zhangf@mail.buct.edu.cn}

\author[mymainaddress]{Chen Hu}
\author[mymainaddress]{Qiang Yin}
\author[mymainaddress]{Wei Hu}

\address[mymainaddress]{College of Information Science and Technology, Beijing University of Chemical Technology, Beijing $100029$, China}

\begin{abstract}
FFT (fast Fourier transform) plays a very important role in many fields, such as digital signal processing, digital image processing and so on. However, in application, FFT becomes a factor of affecting the processing efficiency, especially in remote sensing, which large amounts of data need to be processed with FFT. So shortening the FFT computation time is particularly important. GPU (Graphics Processing Unit) has been used in many common areas and its acceleration effect is very obvious compared with CPU (Central Processing Unit) platform. In this paper, we present a new parallel method to execute FFT on GPU. Based on GPU storage system and hardware processing pipeline, we improve the way of data storage. We divided the data into parts reasonably according the size of data to make full use of the characteristics of the GPU. We propose the memory optimized method based on share memory and texture memory to reduce the number of global memory access to achieve better efficiency. The results show that the GPU-based memory optimized FFT implementation not only can increase over 100\% than FFTW library in CPU platform, but also can improve over 30\% than CUFFT library in GPU platform.
\end{abstract}


\maketitle

\section{Introduction}
FFT is a fast algorithm of DFT (Discrete Fourier transform), it is presented by Cooley and Tukey. The number of multiplications that computer needs to calculate is greatly reduced. But in some areas, the traditional serial operation still can't satisfy requirement. As development of parallel computing, parallel operation of FFT algorithm in many areas plays a very important role. GPU is known as graphics card, it is mainly used for processing graphs. It has powerful computation ability, high-bandwidth and low power consumption characteristics. Because of these, it has been used in many general fields. In the past we have to mapping the algorithm for graphic language to process in the graphics card. CUDA (Compute Unified Device Architecture) is a new kind of technology which is proposed by NVIDIA. Developers can now use C language to write a program for CUDA architecture and not need the knowledge of graphics. We can do parallel programming in GPU by CUDA. It has developed to its fifth generation. Base on the CUDA, the main architecture is CPU + GPU heterogeneous hybrid parallel computing. This heterogeneous platform is gradually applied in a variety of fields.Logic complicated operation will be done in CPU and Parallel part will be finished in GPU.This heterogeneous platform can obviously shorten the calculation time in many fields.

The researchers put forward many methods of FFT acceleration. Yuri Dotsenko showed us that usage of different memory to achieve its optimization. They compute FFT in registers and exchange the data through the share memory [1]. Zhao Li-li adopted a new storage mode of twiddle factor in Texture Memory to improve its performance. This method has a certain accelerating effect for the fixed data. However, it is not very useful for variable data [2]. S. Mitra and A. Srinivasan presented how to accelerate DFT of small amounts of data by GPU. They do the calculations and transpose on the GPU then put the matrix-vector into the constant memory and put the data into the share memory [3]. Sara S. Baghsorkhi presented a solution of providing performance feedback for highly multithreaded graphics processors from the GPU memory hierarchy [4]. Shuai Che proposed the framework though layout remapping and index transformation to improve the efficiency of memory accesses [5]. These give us lots of inspiration.

In this paper, we introduce how to make full use of GPU to accelerate FFT algorithm from the point of data access. Firstly, we will calculate the sine and cosine function according to certain angle. We put the real part and the imaginary part of twiddle factor into the texture memory so that it is convenient to find in the operation. Secondly, the storage of the global memory is large and all the blocks can access in one kernel function. However, the speed of its access is slow. We take advantage of its characteristics and data partitioning and integration are completed in the global memory. Finally, the different blocks of the data are respectively calculated in the share memory. Because the access speed of share memory is faster than global memory. We reduce the use of the global memory and increase the use of share memory compared to the previous method. The number of the global memory access significantly reduces and all the FFT calculation is completed in the share memory. Using this method can save a lot of access time. In order to ensure the coalesced access, we will allocate thread reasonably. There will be no bank conflict when we use share memory. Then we will discuss the details of our implementation and then show the experimental result. And we made a contrast with CUFFT and FFTW. Finally we make a conclusion of this method.

\begin{figure*}[!t]
\centering
\includegraphics[width=120mm]{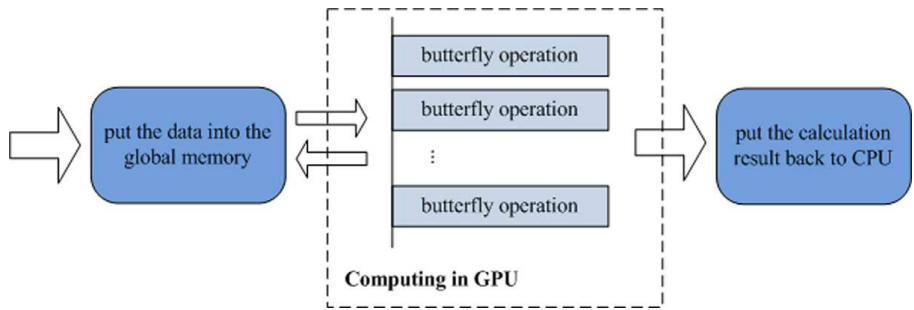}
\caption{calculation process}
\end{figure*}

\section{Our Approach}
In this part, we will show our approach and discuss the detail of our implementation. In section 1, we will focus on the FFT algorithm. In section 2, we will mainly introduce the previous method of FFT algorithm. In the last section, we will show our improved method and analyze the feasibility of this approach theoretically.

\subsection{FFT Algorithm}
DFT (Discrete Fourier Transform) is a form that CFT (Continuous Fourier transform) is discrete both in time domain and frequency domain. DFT has a wide range of applications in many areas. It is suitable for computer processing. But it has large amount of calculation. Its transformation and inverse transformation are expressed as follows:

\begin{equation}
\begin{array}{l}
 X(k) = \sum\limits_{n = 0}^{N - 1} {x(n)W_N^{nk} } , \\
 {\rm{     }}k = 0,1, \cdots ,N - 1 \\
 \end{array}
\end{equation}

\begin{equation}
\begin{array}{l}
 x(n) = \frac{1}{N}\sum\limits_{k = 0}^{N - 1} {X(k)W_N^{ - nk} ,{\rm{     }}}  \\
 {\rm{     }}n = 0,1, \cdots ,N - 1 \\
 \end{array}
\end{equation}

The differences are only in the index of the symbol and the factor, both of them have the same amount of computation. And coefficient has periodic, symmetrical and reducible characteristics. These characteristics are as follows:

\begin{equation}
\begin{array}{l}
 W_N^{nk}  = W_N^{n(k + N)} ,W_N^{nk}  = W_N^{(n + N)k}
\end{array}
\end{equation}

\begin{equation}
\begin{array}{l}
 (W_N^{nk} )^*  = W_N^{ - nk}
\end{array}
\end{equation}

\begin{equation}
\begin{array}{l}
 W_N^{nk}  = W_{mN}^{{\rm{mn}}k} ,W_N^{nk}  = W_{N{\rm{/}}m}^{{\rm{n}}k{\rm{/}}m}
\end{array}
\end{equation}

According to the DFT feathers, researchers improved the DFT algorithm and put forward the FFT algorithm. It improved a big step for the DFT application in the computer. FFT algorithm greatly reduces the computation of the computer and save the storage unit. The calculation time is greatly shortened. It plays a very important significance in many fields. The FFT arithmetic is usually showed by the butterfly operation diagram (see Fig.1).

\begin{figure}[H]
\centering
\includegraphics[width=75mm]{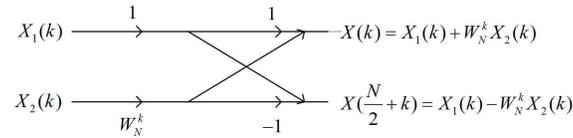}
\caption{butterfly operation}
\end{figure}

\subsection{Previous Method}
We usually use CPU to do the FFT operation without parallel computing before. A butterfly operation is done each time. This traditional method will spend a lot of time when the data volume is large. With the development of parallel computing, some researchers realized the FFT through MPI or some special processing chip like DSP and FPGA. But their cost are too high, using them to do FFT is not the most appropriate in some areas. The launch of CUDA makes many researchers focus on it.

The storage system of GPU is very different from CPU. Its memory is composed of register, local memory, share memory, global memory, constant memory and texture memory. In the Fermi architecture, it adds the cache mechanism. Access speed of share memory is faster than global memory.

The traditional method by GPU is that doing parallel computing each level. The data will be put into global memory. The kernel function will complete a computation of every level. To insure the kernel function being done we need to call many times. Then the next kernel function is called to do the operation of next level. (Fig.2).

\subsection{Our Improved Method}

In this section, we subdivide our improved method into three parts. Firstly, we will show the use of texture memory. Secondly, we will show the access mode optimization. Finally, we introduced our thread allocation.

\begin{figure*}[!t]
\centering
\includegraphics[width=150mm]{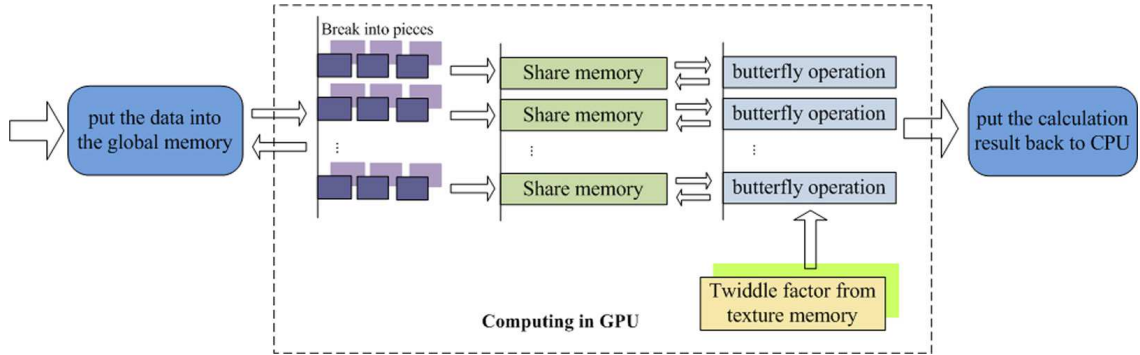}
\caption{Flowchart  of this algorithm}
\end{figure*}

\subsubsection{The use of texture memory}
CPU can only read the data in the internal memory and CPU cache. All the computer programs are running in the internal memory. Access speed of internal memory is relatively slow. It increases caching mechanism to speed up. It is located between the CPU and memory and its access speed is faster than the internal memory. It has a good acceleration effect for the data that CPU often used.
Storage structure has changed a lot in GPU. Share memory and register is small but fast. They belong to the memory chip. Normally their access delay is very small. Other types of memory are located in video memory. Access delay of them is lager and requires 400-600 cycles usually. When we do the calculation by GPU, We should try our best to avoid using the global memory and use the share memory to replace. Access speed of texture memory and constant memory is fast than global memory. These two kinds of memory are suitable for some special treatment. The largest in these memories is global memory but it is the slowest. We draw histogram of their bandwidth and storage size to display intuitively. (Fig.3).

\begin{figure}[H]
\centering
\includegraphics[width=80mm]{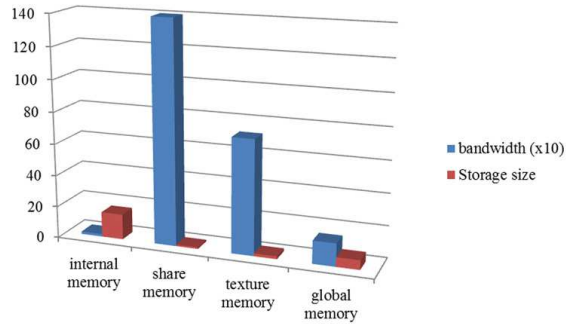}
\caption{histogram of bandwidth and storage size}
\end{figure}

The access speed of texture memory is faster than global memory and slower than the share memory. The data in the texture memory is suitable for look-up table. Before we do the FFT computing, we firstly calculate the value of sine and cosine according the segmentation by certain angle. And then we put the calculated data into the texture memory. When we need the twiddle factor in the butterfly operation, we can query from the texture memory. The method will reduce our operation time to a certain extent.

\subsubsection{Access mode optimization}
As we can see in section 2.2, when the data volume is large, the number of threads accesses global memory will increase. It will spend some extra time. So we think that the share memory will be better and it has a function to ensure synchronous. The storage space of share memory is limited. So in this section we will introduce our solution. In order to make full use of share memory and reduce the use of global memory, we make a big change in accessing data. The execution unit of GPU is called warp. Each warp has 32 threads. So when visiting and fetching data, we do the corresponding change based on the GPU execution to reach the optimum performance.

Our method is as follows. We introduce an example of two times exchange between share memory and global memory. If the data was larger, we will choose more times exchange.

\begin{itemize}
\item
Prepare the data that need to do the FFT opration in CPU.
\item
Put the data into the global memory.
\item
Divide the data into several pieces. Then choose the appropriate number of threads in one block. This step we just use its first dimension.

\item
Put the data in different pieces into share memory. And do the butterfly operation in each block. This step there is no bank conflict in share memory and the previous level will be done.

\item
Put the data in share memory back into the global memory. This is the first time exchange.

\item
Continue to divide the pervious data into several pieces. Then make the small pieces together. Choose the number of threads in one block. This step will make sure that first dimension is 16. Because the coalescent is needed. Do the butterfly operation according to the column. There is also no bank conflict in this step.

\item
Put the data in different pieces into share memory and do the butterfly operation in each block. In this step the later level will be done. And all the butterfly operation will be completed. In this step the later level will be done. And all the butterfly operation will be completed.

\item
Put the data in share memory back to the global memory. This is the last time exchange.
\item
Put the calculation results in global memory back to CPU.
(see Fig.4).
\end{itemize}

\begin{figure}[!hbp]
\centering
\includegraphics[width=80mm]{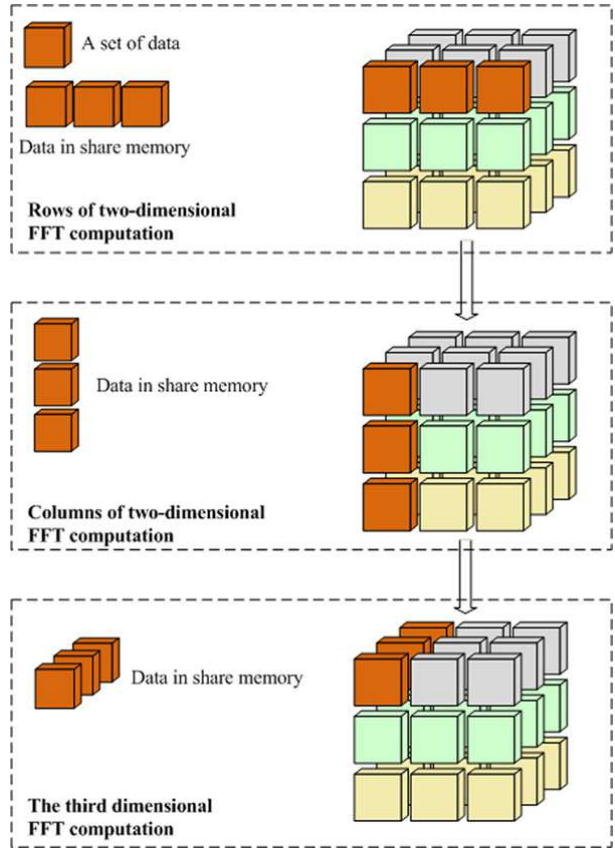}
\caption{the storage of the data}
\end{figure}

When the data quantity is large, we will use the share memory several times. We show an example that we use the share memory three times. (Fig.5). We treat the data stored in the linear as the three-dimensional. Each small cube represents a set of data. We allocate the appropriate number of threads. In order to ensure the combined memory access, the access address of each 32 threads is continuous.

Because of the size limit of the share memory, the previous method is put the data into the global memory. But in this way, a lot of time will be spend on the data access. During the FFT computation, we find that the differences of the butterfly operation in each level are the span and twiddle factor. So we break the data into pieces to make its span small. And every piece is suitable for computing in the share memory. Using share memory is faster than using global memory.



We divide the data into different parts according to the size of the data. When the data quantity is less than 1024, we don't need to divide. If the data quantity is more than 1024 less than 65536 we divided the data into two parts. When the data quantity is larger, we will be divided into more parts. We do that based on the size of the share memory because the size of the share memory is certain. In this method each level of the butterfly operation don't need to access the global memory compared to the previous method.

First we put the piece of data into share memory and do the butterfly operation of the rows. And then we do the butterfly operation of the columns and the third dimensional. The operation in share memory has no bank conflict. We do the exchange twice in global memory and all the butterfly operation is done in the share memory. So we can reduce the use of the global memory and save the time of data access.

\subsubsection{Threads allocation}
The kernel function executes on the GPU. We need allocate a reasonable number of threads to ensure the effective implementation of the kernel function. Each kernel usually has a thread grid. It consists of thread blocks. Thread block consists of a number of threads. Block is executed in parallel. Threads in the same block can communicate by the share memory. Different block cannot communicate with each other.

The computing core of GPU is SM (Stream Multiprocessor). Eight SP (Stream Processor) and a small number of other computing units make up one SM. The execution unit of the kernel function is block. In the actual operation, block is divided into a small number of threads called warp. Each warp consists of 32 threads. Eight SP will execute 4 times in one instruction.

Coalesced access can achieve the best effect in global memory. That is to say, Continuous thread needs access to the continuous address. In the Fermi architecture, the 32 threads in one warp will be parallel execution by an instruction. In order to ensure the coalesced access, we need to make sure that the access address of 32 threads in one warp is continuous. The data will be divided into three-dimensional as in the section 2.3.2. The three dimension size of the block is set to 32, 16, 1. The size of the grid will be decided according to the data size. The rows exchange the data between share memory and global memory and the columns do the butterfly operation.

When the data store in the share memory, the best effect will be achieved without bank conflict. There are usually 16 banks in each share memory. If multiple threads access the same bank, there will be bank conflict. This will greatly reduce operation efficiency. However, the bank will broadcast and not affect the access speed when the half-warp access the same bank. The size of first dimension that we set is 16 and the size of second dimension is 33. When the threads in half-warp access the share memory there will be no bank conflict.

\begin{figure}[!hbp]
\centering
\includegraphics[scale=0.45]{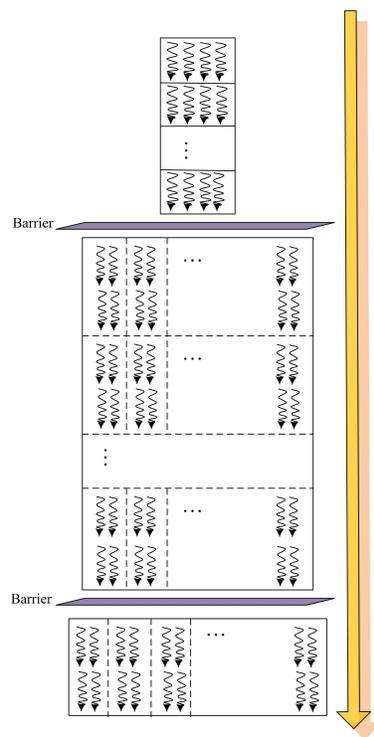}
\caption{Threads allocation}
\end{figure}

\begin{table*}[!tb]
\center
\caption{Comparison of efficiency}
\begin{tabular}{cccc}
\hline
{\bf Application} & {\bf FFTW (ms)} & {\bf CUFFT (ms)} & {\bf Our FFT (ms)} \\
\hline
16&0.015377&0.344384& 0.170848 \\

\hline
64& 0.029687&0.358176&0.178016\\

\hline
256& 0.050903&0.350688& 0.180192\\

\hline
1024& 0.043384&0.405088&0.194880\\

\hline
4096&0.120041&0.416288&0.208768\\

\hline
16384& 0.428061&0.504672&0.294368\\

\hline
65536&1.489800&0.91008&0.792608\\
\hline
\hline

%
%
%
%
%
%
%
%
%

\end{tabular}
\end{table*}

As we can see in Fig.6, each thread calculates a butterfly operation. The fence can guarantee the synchronization of each block. We call the kernel function three times to make sure synchronization when we use global memory.The three parts in Fig.6 correspond to the three dimensions in Fig.5.

\section{Experimental Results}
In this part, we carry out FFT computing experiment and made a contrast with FFTW (the Faster Transform in the West) and CUFFT (official FFT library on GPU). Our experiment environment is composed of Tesla C2070 GPU and intel (R) Core (TM) i7-2600K CPU. Due to share memory limitation, we will call the kernel function to calculate once when the data volume is less than 1024. If the data volume is more than 1024 and less than 32768, we will call the kernel function twice. When the data volume is larger, we will call the kernel function more times.

\begin{figure}[h]
\centering
\includegraphics[width=80mm]{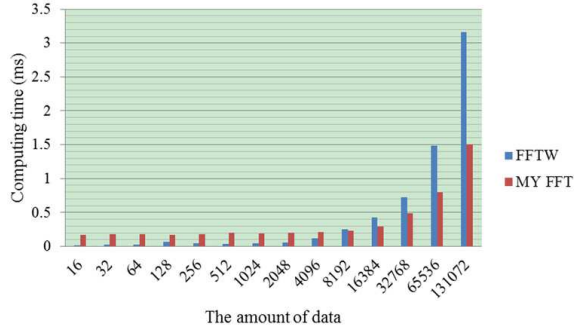}
\caption{Speed comparison with FFTW}
\end{figure}

\begin{figure}[h]
\centering
\includegraphics[width=80mm]{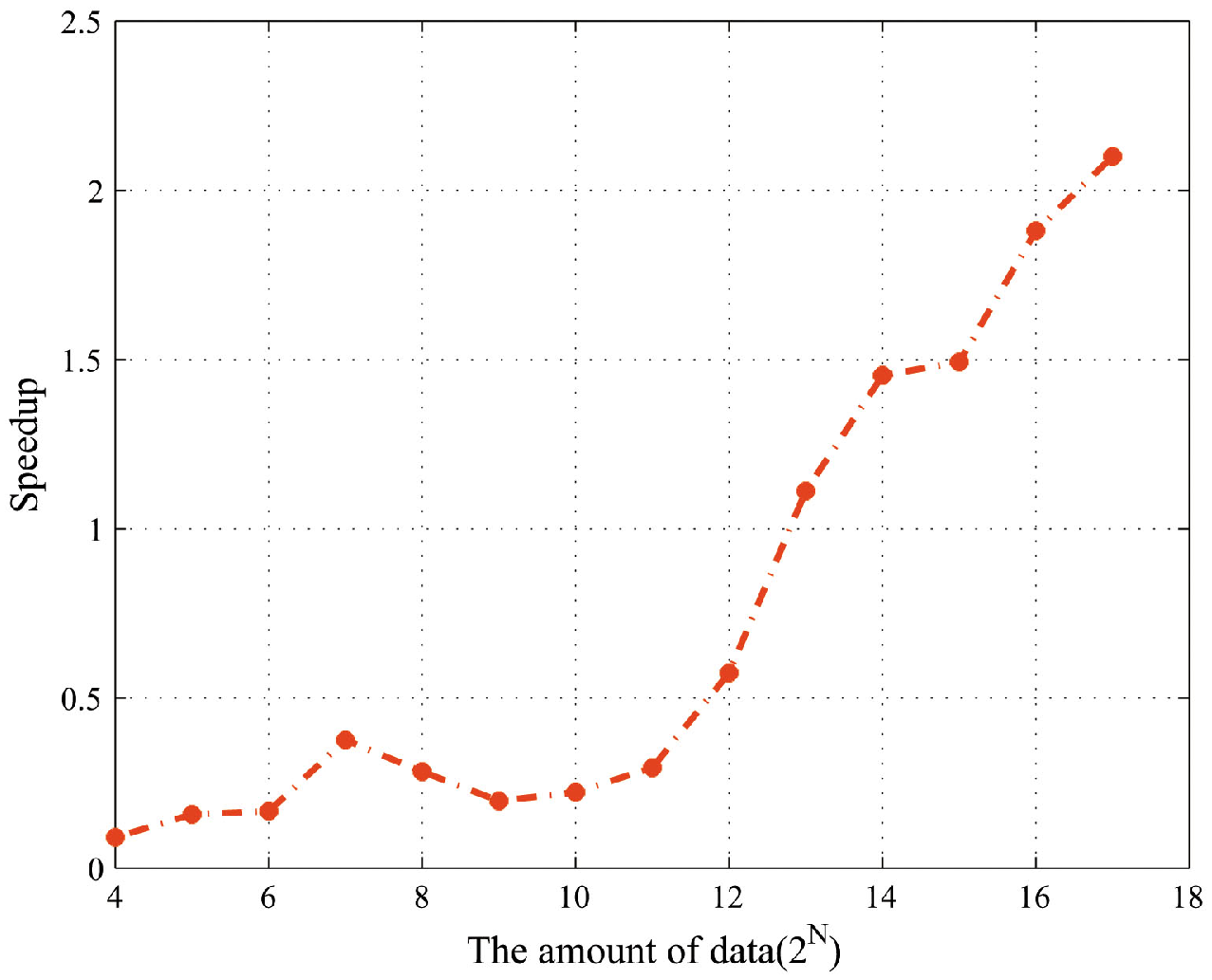}
\caption{Speed comparison with FFTW}
\end{figure}

FFTW is standard C language assembly of FFT. It is usually faster than the other open source FFT program. The result that contrasts with FFTW is shown above. From Fig.7-8 we can see that the speed of FFTW is faster when the data volume is less than 8192. In our FFT, the data need to be transfered from internal memory to video memory with GPU computing, so when the data volume is small, most of the time consumed in the data transmission. When the data volume is less than 4096, the curve is relatively stable. With the growth of the data volume, accelerating effect appears gradually. When the data volume is 65536, we called the kernel function three times. So the accelerating effect has a certain decline. From the graph, we can see that the accelerating effect is gradually obvious as a whole with the increase of the data volume.

\begin{figure}[H]
\centering
\includegraphics[width=80mm]{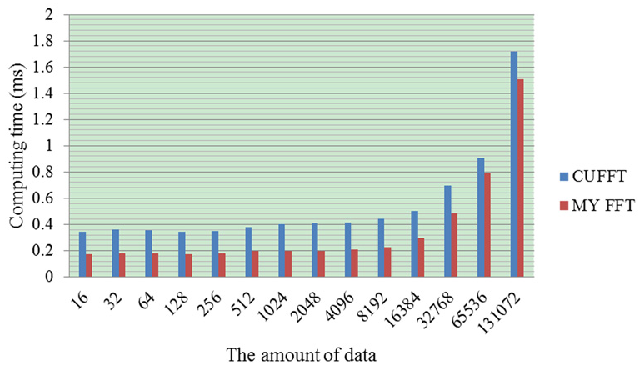}
\caption{Speed comparison with FFTW}
\end{figure}

\begin{figure}[H]
\centering
\includegraphics[width=80mm]{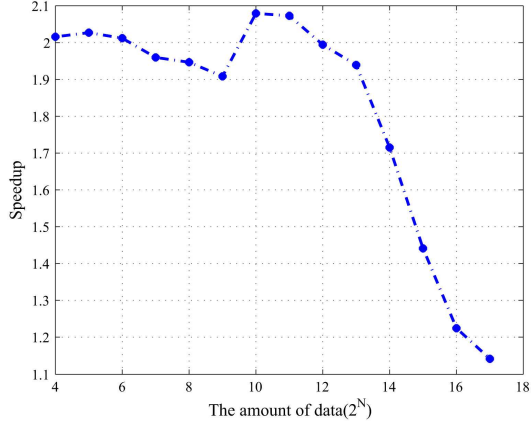}
\caption{Speed comparison with CUFFT}
\end{figure}

 As show in Fig.9-10, out FFT is better than CUFFT after the memory optimization. The acceleration effect is more obvious in moderate length signals. Due the the limitation of share memory, the speedup will decrease with the increase of signal length. And within a certain range of data, there is also a certain degree of improvement compared to CUFFT. In the SAR (synthetic aperture radar) imaging processing, the data scale of FFT operation is from a few thousands to tens of thousands. There is a big improvement about 30\% to 100\% compared to the previous. Thus the optimized FFT implementation will benefit the GPU-base SAR processing algorithms a lot.

From the experiments, it is clear that FFT operation is also faster to calculate with CUFFT by NVIDIA. The reason may be that these operations are processed in the unit of SFU in GPU. Based on the Fermi architecture, it increases cache mechanism. So a few number of accessing global memory spends a little time compared with the large amounts of calculation. we also make a contrast of accuracy between CPU and GPU. Our method meets the requirement of accuracy. From results, we can conclude that GPU has no advantage when the sequence is small, where FFTW is dominant. However, as the increase of data volume, parallel advantage of GPU appears gradually.

\section{Conclusion}
In this paper, we introduce a new method to improve the FFT parallel performance on GPU from the point of memory access. We reduce the number of global memory access and make full use of high bandwidth characteristics of share and texture memory to achieve the improvement of acceleration effect. Through the acceleration of FFT, some FFT-based algorithms can achieve better computing efficiency,such as SAR simulation, imaging and so on. So optimizing FFT operation is of great significance. GPU computing still has its bottleneck at the data transfer, which consumes a lot of time in data transfer from CPU to GPU . We will continue to improve our method from the data transmission to get the better acceleration in our next research.


\end{document}